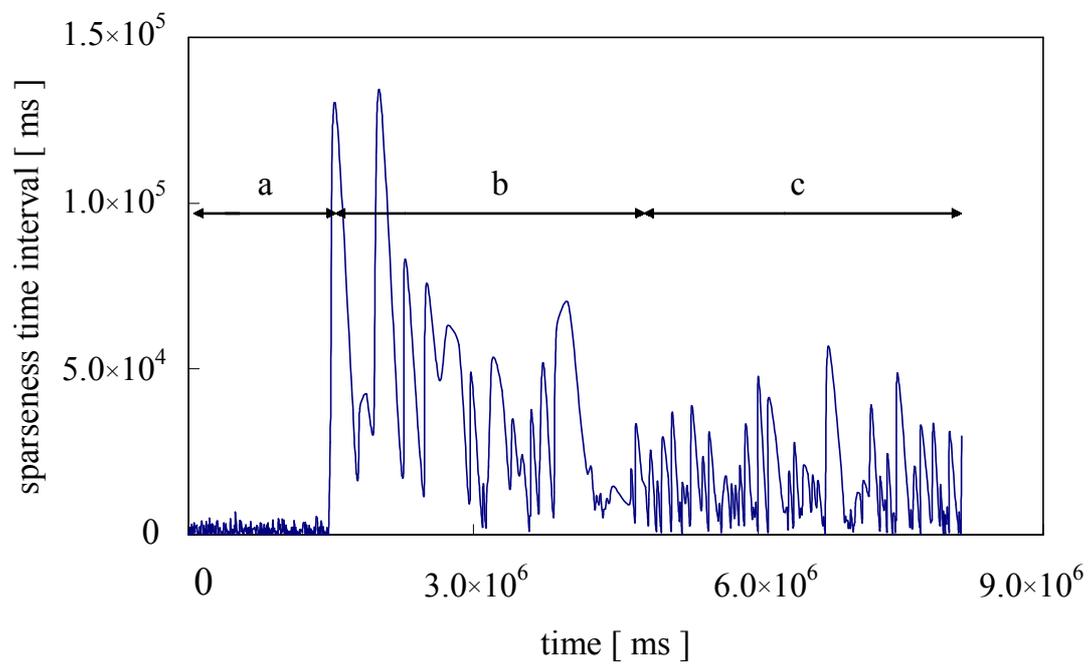

Fig. 1

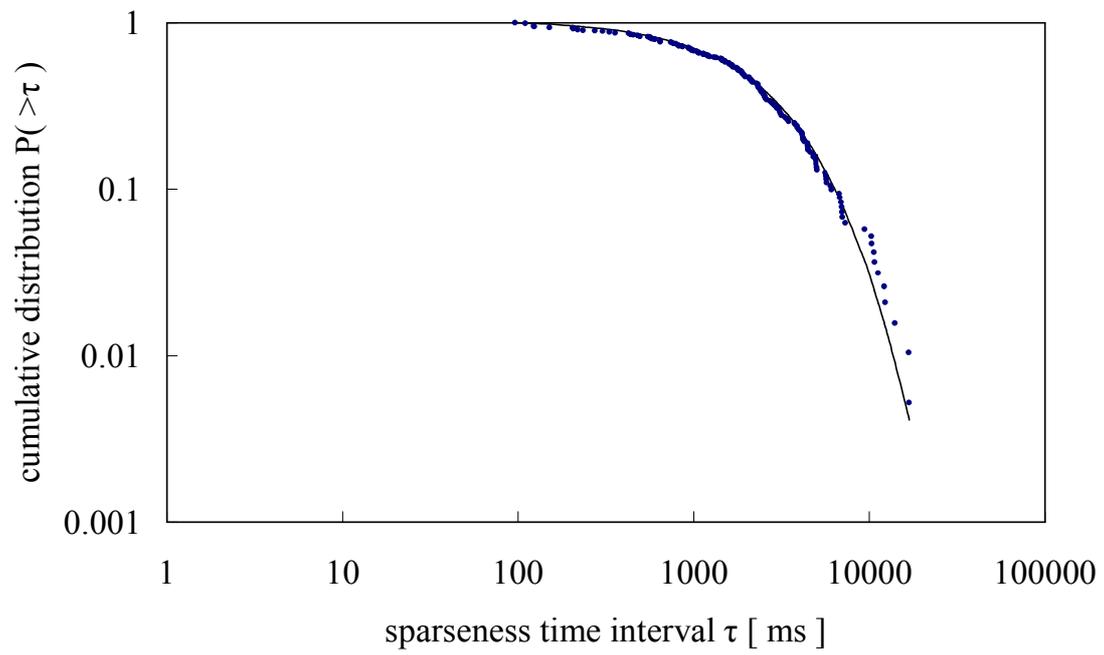

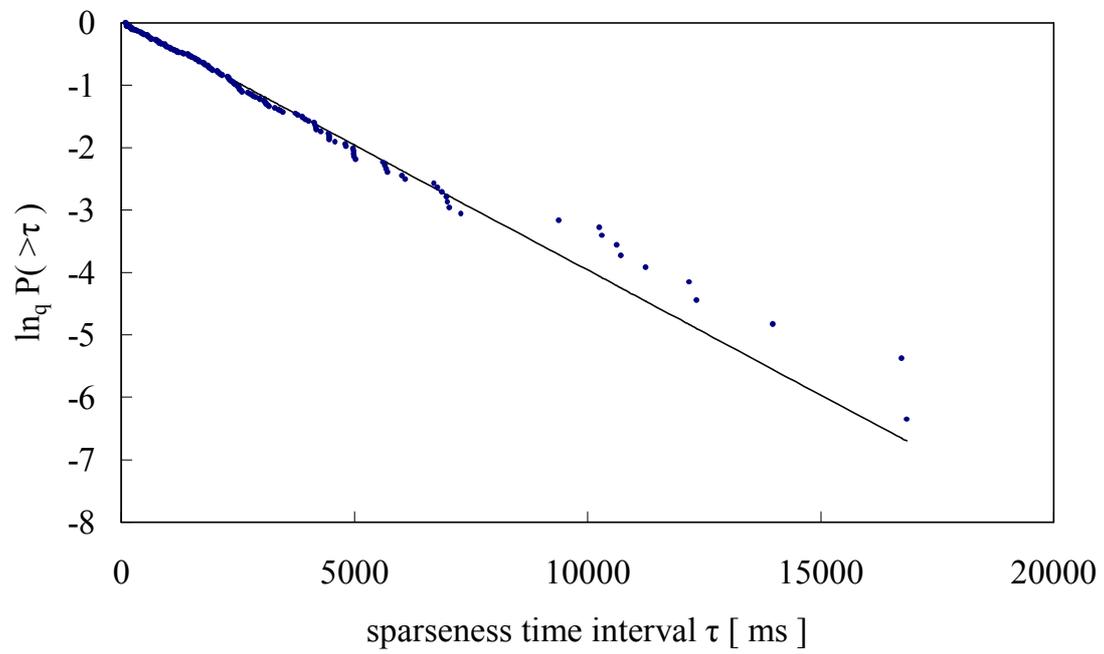

Fig. 2a

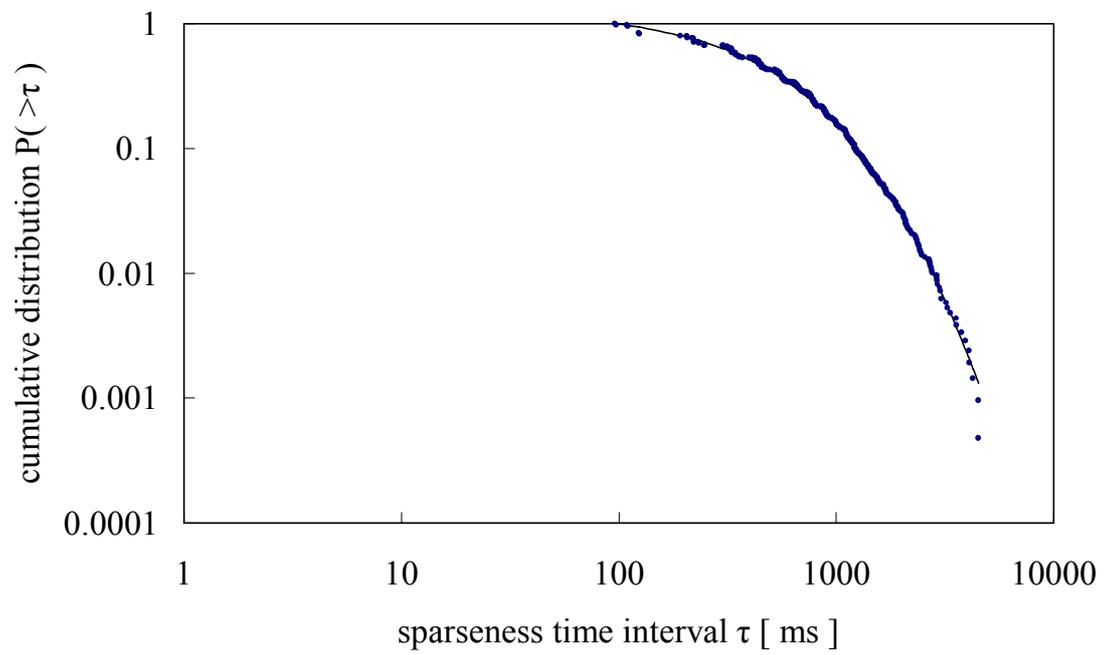

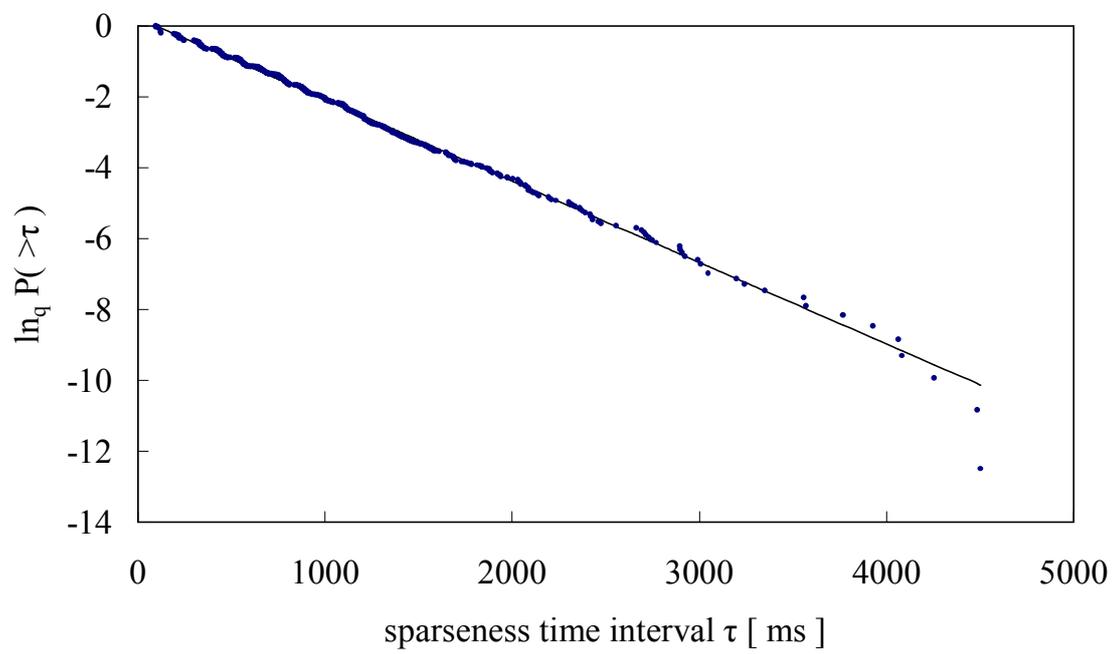

Fig. 2b

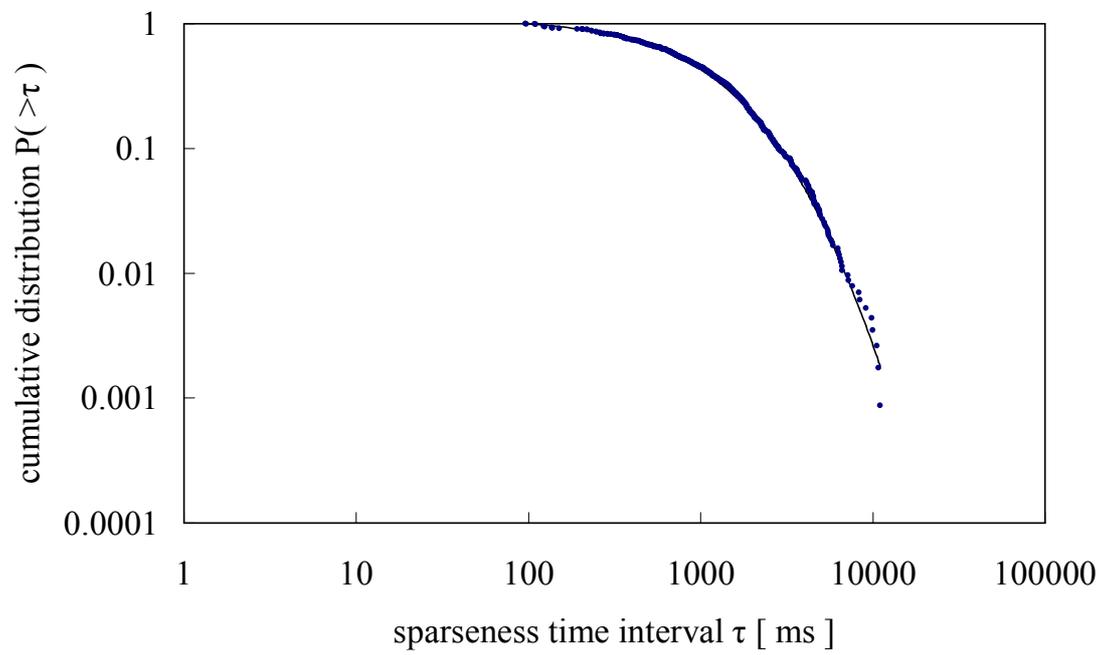

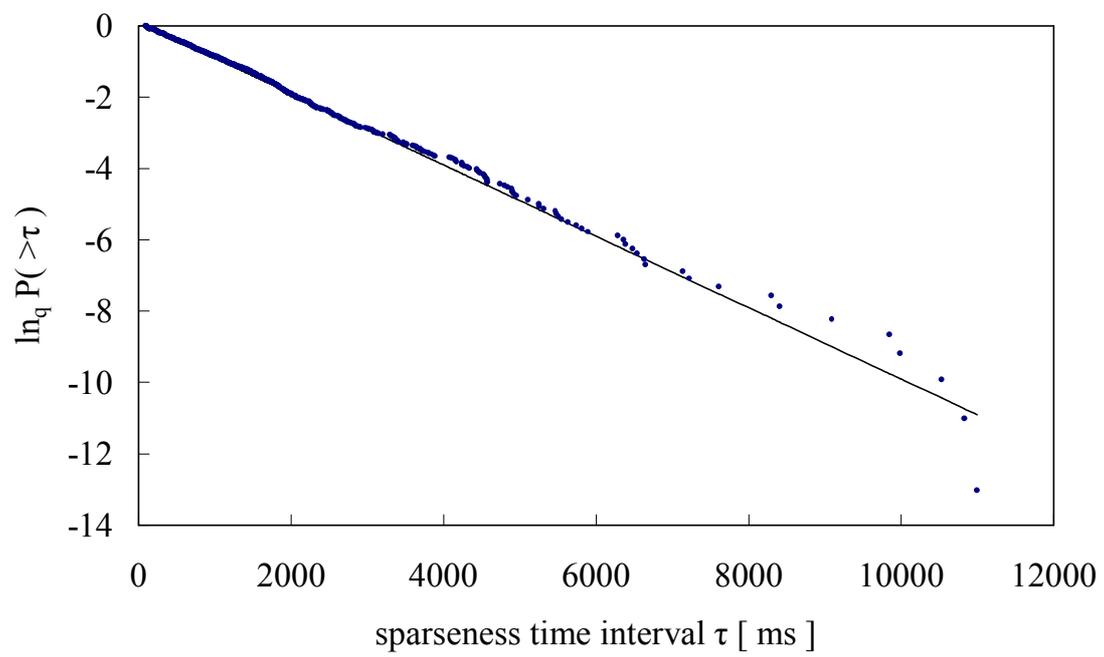

Fig. 2c

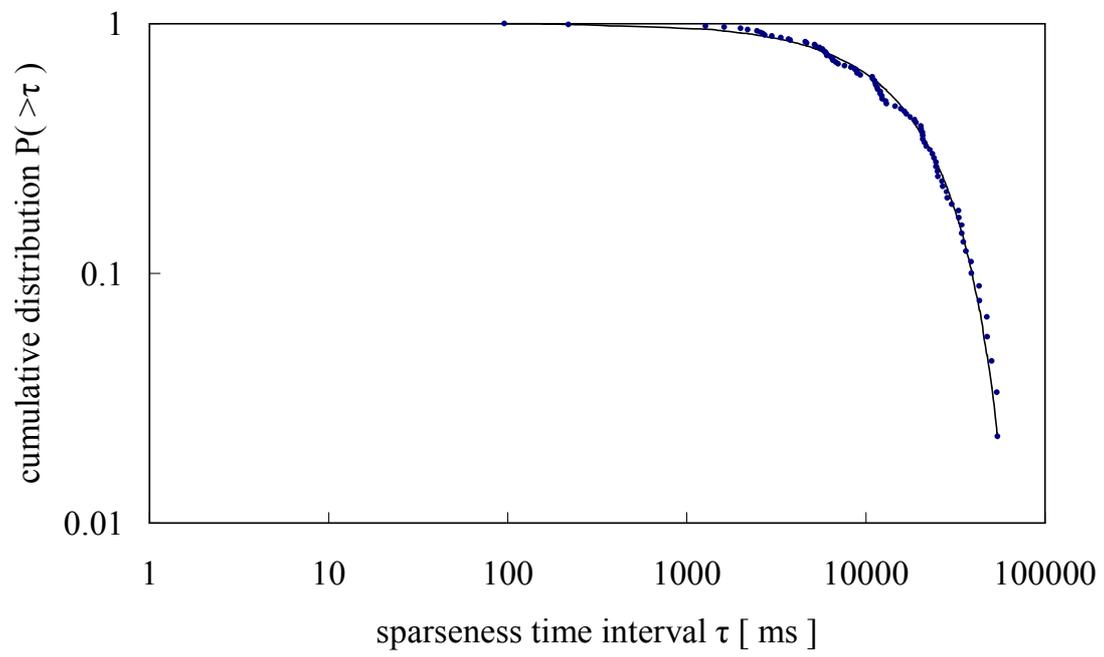

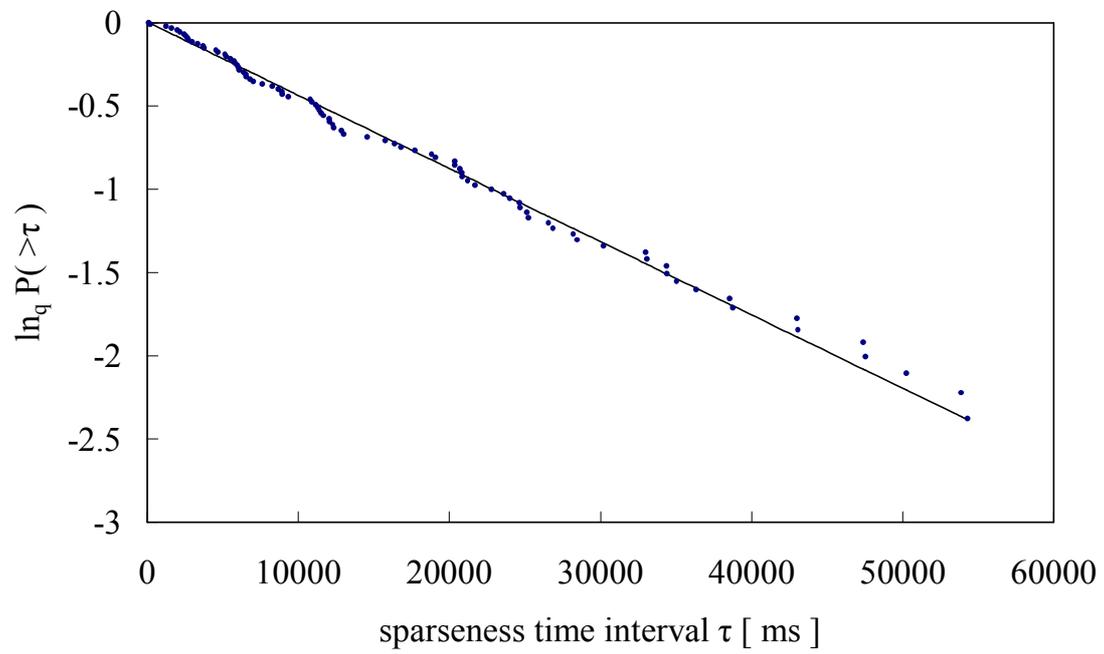

Fig. 3

# Itineration of the Internet over nonequilibrium stationary states in Tsallis statistics

Sumiyoshi Abe[1] and Norikazu Suzuki[2]

[1]*Institute of Physics, University of Tsukuba, Ibaraki 305-8571, Japan*

[2]*College of Science and Technology, Nihon University,*

*Funabashi, Chiba 274-8501, Japan*

The cumulative probability distribution of sparseness time interval in the Internet is studied by the method of data analysis. Round-trip time between a local host and a destination host through ten odd routers is measured using the Ping Command, i.e., doing echo experiment. The data are found to be well described by the $q$-exponential distributions, which maximize the Tsallis entropy indexed by $q$ less or larger than unity, showing a scale-invariant feature of the system. The network is observed to itinerate over a series of the nonequilibrium stationary states characterized by Tsallis statistics.

The Internet is a complex system, which has highly intricate tangle, cluster and hierarchical structures, strong spatio-temporal correlation with feedback, self-organization, and connection diversity. The structure emerging from the actions of a large number of the users may efficiently be understood within the statistical mechanical framework and its suitable generalizations. For example, a work in [1] reports emergence of scaling behavior and the associated power-law distribution of connectivity of nodes. These concepts are known to be essential for the network to be resilient and robust to random errors, breakdown, and attack [2-4].

From the statistical and dynamical viewpoints of the network, of particular interest are the stationary states under nonequilibrium conditions. Tsallis statistics [5] based on a nonextensive entropy [6] aims to offer a theoretical basis for analyzing complex systems at such states. It has successfully been applied to a variety of problems including anomalous diffusion [7,8], Lévy flight [9-11], fractal random walk [12], complex high-energy processes [13-17], cosmic rays [18], turbulence [19], earthquakes [20], stock markets and incomes [21,22], nonlinear maps at the edge of chaos [23-29], stochastic resonance [30], protein folding and biomolecules [31,32], citation networks of scientific papers [33], urban agglomeration [34], and linguistics [35].

In this Letter, we present an experimental evidence that Tsallis statistics in fact describes the scale-invariant stationary states of the Internet.

"Echo experiment" we have performed uses the Ping Command [36,37]. A Ping signal is emitted from a local host computer, takes a round trip to a destination host (i.e., a site accessed), and returns to the local host through ten odd routers. The route of the signal emitted to the destination is fixed and traced. Each router is connected with the whole network in a time-dependent manner. The next signal is sent immediately after the previous one returns. Such a time interval is typically $<130\,\text{ms}$ and is not included in our data analysis. Using the whole collected data of the echo experiment, the threshold value indicating congestion is appropriately defined (see the later discussion). Actually, *the result turned out to be not sensitive to the definition of the threshold*. We have calculated each time interval given by the amount of the round-trip time below the threshold value between two successive thresholds. This interval is referred to here as "sparseness time interval" denoted by $\tau$. Observation shows that the Internet itinerates over a series of the stationary states, which are all described by the Tsallis *q*-exponential cumulative probability distribution of $\tau$, indicating a scale-invariant feature of the system. In particular, both the $0<q<1$ and $q>1$ cases occur and nothing is special in the limit $q \to 1$. Regarding sparseness time interval rescaled by its average, the more congested the network is, the smaller value $q$ takes. In this sense, the entropic index characterizes the degree of congestion.

Before presenting the experimental results, let us briefly summarize the basics of Tsallis' scale-invariant statistics. This theory aims to offer a framework for describing statistical properties of complex systems at their stationary states based on principle of maximum nonextensive entropy. In the present case, the fundamental random variable is sparseness time interval, $\tau$. $p(\tau)\,d\tau$ is the probability of finding the value of sparseness time interval in the range, $[\tau, \tau+d\tau]$. Then, $p(\tau)$ is a stationary probability distribution in Tsallis statistics if it optimizes the Tsallis entropy [6,38]

$$S_q = \frac{1}{1-q}\left(\int \frac{d\tau}{\sigma}[\sigma p(\tau)]^q - 1\right) \tag{1}$$

under the constraints on normalization

$$\int d\tau\, p(\tau) = 1 \tag{2}$$

and the normalized $q$-expectation value [39] of sparseness time interval

$$<\tau>_q = \int d\tau\, \tau\, P_q(\tau). \tag{3}$$

Here, $q$ and $\sigma$ in Eq. (1) are the positive entropic index and a scale factor of the dimension of time, respectively. $P_q(\tau)$ in Eq. (3) is the escort distribution [40] defined by

$$P_q(\tau) = \frac{p^q(\tau)}{\int d\tau'\, p^q(\tau')}. \tag{4}$$

The optimal distribution is calculated to be

$$p(\tau) = \frac{1}{Z_q} e_q\left(-\frac{\beta}{c}(\tau - <\tau>_q)\right), \tag{5}$$

$$Z_q = \int_0^{\tau_{max}} d\tau \, e_q\left(-\frac{\beta}{c}(\tau - <\tau>_q)\right), \tag{6}$$

$$c = \int_0^{\tau_{max}} \frac{d\tau}{\sigma}[\sigma p(\tau)]^q. \tag{7}$$

$\beta$ in Eqs. (5) and (6) is the Lagrange multiplier associated with the constraint in Eq. (3). $e_q(x)$ stands for the "q-exponential function" defined by

$$e_q(x) = \begin{cases} [1+(1-q)x]^{1/(1-q)} & (1+(1-q)x \geq 0) \\ 0 & (1+(1-q)x < 0) \end{cases}, \tag{8}$$

whose inverse is the "q-logarithmic function"

$$\ln_q(x) = \frac{x^{1-q} - 1}{1-q}. \tag{9}$$

Accordingly, $\tau_{max} \to \infty$ if $q \geq 1$, whereas $\tau_{max} = \tau_0/(1-q)$ if $0 < q < 1$. Here, $\tau_0 = [c+(1-q)\beta <\tau>_q]/\beta$, which can be shown to be always positive [11]. $p(\tau)$ is recast into the following form:

$$p(\tau) = \frac{e_q(-\tau/\tau_0)}{\int_0^{\tau_{max}} d\tau' e_q(-\tau'/\tau_0)}. \tag{10}$$

$p(\tau)$ is seen to be the Zipf-Mandelbrot distribution with a heavy tail if $q > 1$. Both the normalizability condition and finiteness of $<\tau>_q$ in Eq. (3) require the entropic index to satisfy $q < 2$.

In the limit $q \to 1$, the Tsallis entropy converges to the Boltzmann-Shannon entropy, $S = -\int d\tau\, p(\tau) \ln[\sigma\, p(\tau)]$, and correspondingly $p(\tau)$ becomes the Boltzmann-type exponential distribution since, in this limit, $e_q(x)$ and $\ln_q(x)$ approach to the ordinary exponential and logarithmic functions, respectively. However, such a limit does not play any special roles in the present work.

An important point in Tsallis statistics is that the quantity to be compared with the observed distribution is not $p(\tau)$ in Eq. (10) itself but its associated escort distribution [41]. Therefore, the cumulative probability distribution should be defined by $P(>\tau) = \int_\tau^{\tau_{max}} d\tau'\, P_q(\tau')$. From Eq. (10), it is found to be given by

$$P(>\tau) = e_q(-\tau/\tau_0). \qquad (11)$$

Below, we discuss how the cumulative probability distributions of this form are realized in the Internet.

In Fig. 1, we present an example of an observed time series of sparseness time interval, $\tau$. Three distinct stationary regimes, a, b, and c, may be recognized. (Strictly speaking, identification of stationary state depends on time scale. Here, we are employing user's typical time scale, i.e., 10 minutes ~ 1 hour.) In Fig. 2 a, b, and c, the corresponding cumulative probability distributions of $\tau$ are plotted on the log-log scale. The threshold value indicating congestion is defined here by the mean value plus half of the standard deviation. The experimental data are represented by the dots, whereas the curves depict the *q*-exponential functions. In particular, the lower ones are drawn on the "semi-*q*-log" scale with different values of *q*. The resultant straight lines imply that the observed cumulative probability distributions are in fact the Tsallis *q*-exponential distributions.

For comparison, we present Fig. 3 to show that *there also exist stationary states, at which the values of the entropic index are less than unity*.

These results imply that the network undergoes a series of transition from one stationary state to another: $(q_1, \tau_{0,1}) \to (q_2, \tau_{0,2}) \to (q_3, \tau_{0,3}) \to \mathtt{L}$. Each stationary state is scale-invariant and maximizes the Tsallis entropy. The points of transition correspond to a catastrophic changes in the time series of round-trip time (not sparseness time), e.g., sudden heavy congestion.

In conclusion, we have found that the Internet itinerates over a series of the scale-invariant nonequilibrium stationary states described by Tsallis statistics. We wish to emphasize that the time series of sparseness time is highly nonstationary and non-Gaussian. This fact makes it difficult to identify stationary regimes by power spectrum analysis, in general. The present work indicates usefulness of Tsallis statistics for defining stationary states in the time series exhibited by complex systems.

**Acknowledgment**

We would like to thank Dr. V. Latora and Dr. M. Takayasu for useful discussions.

Figure Captions

Fig. 1    Time series data of the sparseness time interval taken from 3.50 *a.m.* (initial time) to 6.07 *a.m.* on 8 February, 2002. From the local host, buffalo.matsudo-ap3.dti.ne.jp [203.181.67.200], to the destination host, ring.so-net.ne.jp [202.238.95.103], through 11 routers. The curve is drawn based on 31675 measured data points. Roughly, three different nonequilibrium stationary states, **a** (3.50 *a.m.* – 4.15 *a.m.*), **b** (4.15 *a.m.* – 5.06 *a.m.*), and **c** (5.06 *a.m.* – 6.07 *a.m.*), may be recognized.

Fig. 2    Log-log plots of the cumulative probability distributions associated with the the states, **a**, **b**, and **c**. The observed data are represented by the dots, whereas the Tsallis distribution by the solid lines. The lower ones are drawn on the semi-*q*-log scale. **a**: $q = 1.07$, $\tau_0 = 2.50 \times 10^3$ ms, and 4373 data points. **b**: $q = 1.12$, $\tau_0 = 4.35 \times 10^2$ ms, and 13587 data points. **c**: $q = 1.16$, $\tau_0 = 1.00 \times 10^3$ ms, and 13715 data points.

Fig. 3    An example with the value of *q* less than unity. The hosts are the same as in Fig. 1. Data was taken from 4.42 *a.m.* to 5.31 *a.m.* on 13 February, 2002. $q = 0.73$, $\tau_0 = 2.27 \times 10^4$ ms, and 14897 data points.

11